\documentclass[twocolumn]{aa}
\usepackage{graphicx}
\usepackage{txfonts}
\begin{document}
   \title{Confirmation of strong magnetic field amplification and 
nuclear cosmic ray acceleration in SN~1006}

   \author{E. G. Berezhko
          \inst{1}
          \and
          L. T. Ksenofontov\inst{1,2} 
          \and
          H. J. V\"olk\inst{3}
          }

   \offprints{H. J. V\"olk}

   \institute{Institute of Cosmophysical Research and Aeronomy,
                     31 Lenin Ave., 677891 Yakutsk, Russia
         \and
             Institute for Cosmic Ray Research, University of Tokyo,
              Chiba 277-8582, Japan
         \and
             Max Planck Institut f\"ur Kernphysik,
                Postfach 103980, D-69029 Heidelberg, Germany\\
             \email{Heinrich.Voelk@mpi-hd.mpg.de}
             }

   \date{Received  / Accepted }

   \abstract{
It is shown that the nonlinear kinetic model of cosmic ray (CR)  
acceleration in supernova remnants (SNRs) fits the shell-type nonthermal
X-ray morphology, obtained in Chandra observations, in a satisfactory way.
The set of empirical parameters is the same which reproduces the dynamical
properties of the SNR and the spectral characteristics of the emission
produced by CRs. The extremely small spatial scales of the observed X-ray
distribution are due to the large effective magnetic field 
$B_{\mathrm d}\sim 100$~$\mu$G in the interior, which is also required to
give a good fit for the spatially integrated radio and X-ray synchrotron
spectra. The only reasonably thinkable condition for the production of 
such a large
effective field strength is an efficiently accelerated nuclear CR
component. Therefore the Chandra data confirm the inference that SN~1006
indeed accelerates nuclear CRs with the high efficiency required for SNRs
to be considered as the main Galactic CR sources.
   }
   
  \authorrunning{Berezhko et al.} 
  
  \titlerunning{Confirmation of strong magnetic field amplification and 
nuclear CR acceleration in SN~1006}

   \maketitle
%


A simple-minded calculation (McKenzie \& V\"olk \cite{mckv82}) of the  
mean square
magnetic field (MF) fluctuation $(\delta B)^2$
\begin{equation}
(\delta B/B)^2 = M_{\mathrm a} P_{\mathrm c}/(\rho_{\mathrm g}
V_{\mathrm s}^2)
\label{eq1}
\end{equation}
excited by the cosmic ray (CR) streaming instability in a strong
shock with $M_{\mathrm a}$ $\gg$ $1$ shows that for efficient acceleration, $P_{\mathrm c} \sim
\rho_{\mathrm g} V_{\mathrm s}^2$, the excitation of resonant Alfv\'en waves is not only an
important ingredient for the diffusive acceleration mechanism as such
(Bell \cite{bell78}; Blandford \& Ostriker \cite{bo78}). The resultant 
$(\delta B/B)^2 \gg 1$ could 
also
imply a considerable obstacle for the theoretical description of the
process up to the production of an enhanced effective MF in the
acceleration region (V\"olk \cite{voe84}). In
eq.\ref{eq1} $M_{\mathrm a}=V_{\mathrm s}/c_{\mathrm a}$ is the Alfv\'{e}nic Mach number, where
$c_{\mathrm a}$ denotes the Alfv\'{e}n speed, $P_{\mathrm c}$ is the CR pressure at the
shock front, and $\rho_{\mathrm g} V_{\mathrm s}^2$ is the ram pressure of the incoming plasma
flow. The problem of eq.\ref{eq1} was taken up again recently by Bell \&
Lucek (\cite{bluc01}), referring also to (Lucek \& Bell \cite{lucb00}) and
attempting a nonlinear description of the MF evolution in a
simple model, with the conclusion that a considerable amplification to
what we shall call an effective MF should indeed occur. At the
same time the selfconsistent treatment of the nonlinear time-dependent
acceleration equations, both in theoretical and computational terms (e.g.
Berezhko et al.\ \cite{byk96}; Berezhko \& V\"olk \cite{bv97}), had
progressed sufficiently to make quantitative modeling of CR acceleration
in specific supernova remnants SNRs like the historical object SN~1006
possible (Berezhko et al.\ \cite{bkv02}), including the so-called injection
process (V\"olk et al.\ \cite{vbk03} and references therein). This allowed
a detailed comparison with radio and X-ray observations which had become
available in the nineties (Koyama et al.\ \cite{koyama}; Allen et al.\
\cite{apg}), and allowed in particular a theoretical study of the
morphology. In fact there are very recent measurements on details of the
X-ray morphology of SN~1006 with Chandra (Long et al.\ \cite{long}; Bamba
et al.\ \cite{bamba}). We shall show in this Letter that they confirm the
amplification of the MF and - for lack of any acceptable
alternative - its cause, the efficient acceleration of nuclear CRs.

Nonthermal X-ray observations indicate that at least CR electrons are
accelerated in SNRs. In SN~1006 there is evidence that electrons reach
energies of about 100~TeV (Koyama et al. \cite{koyama}; Allen et al.
\cite{apg}). Also TeV $\gamma$-ray emission from this source has been
reported (Tanimori et al. \cite{tan98}). However, depending on the assumed
values for poorly known physical parameters in the theory (mainly the
value of the MF and the nucleon injection rate, but also the
ambient gas density), the observed high-energy $\gamma$-ray emission of
SN~1006 could in principle be predominantly either simply the inverse
Compton radiation of the synchrotron electrons scattering on the microwave
background radiation as assumed by Tanimori et al. (\cite{tan98}), or
$\pi^0 - \,$ decay emission due to hadronic collisions of CRs with gas 
nuclei
(Berezhko et al. \cite{bkv02}). Comparing the synchrotron data with the
calculated spectrum of the energetic electrons Berezhko et al.
(\cite{bkv02}) have inferred that the existing data require a very
efficient acceleration of CR {\it nuclei} at the SNR blast wave (which
converts about 10 \% of the initial SNR energy content into CR energy) as
well as a large interior MF strength $B_{\mathrm d} \approx 120 \mu$G. The large
value of $B_{\mathrm d}$ is expected and possible in this case, using the spatially
integrated radio and X-ray synchrotron spectra to empirically determine
the MF strength and the injection rate in the theory. MF
amplification was also one of the possibilities discussed later by Long et
al. (\cite{long}). The nonthermal energy is distributed between energetic
electrons and protons in the proportion $1.5 \times 10^{-3}$, similar to
that of the Galactic CRs. From the point of view of acceleration theory
this is the physically most plausible solution. It can also explain the
peculiar dipolar structure of the synchrotron emission, with the dipole
axis parallel to the ambient interstellar MF, as an effect of the
homogeneity of the large-scale ambient MF (V\"olk et al.
\cite{vbk03}), generally believed to exist in the rather clean
interstellar environment of SN~1006 above the Galactic disk (Long et al.
\cite{long}). This morphology is fairly common among SNRs (Winkler \& Long
\cite{wink}). There is no alternative process without ad hoc-assumptions 
in the literature, or a new
one which we could reasonably imagine, that would amplify the MF in a
collisionless shock without particle acceleration. Nevertheless, the
previous observations of the {\it large-scale} nonthermal emission in the
case of SN~1006 did not strongly exclude a scenario which we called the
``inefficient model'', with low interior MF $B_{\mathrm d}\approx 10~\mu$G
in which nuclear CRs would play no important role and practically all
nonthermal emissions would be of leptonic origin (Berezhko et al.
\cite{bkv02}). As a corollary we shall show here that such a low-field
scenario is not compatible with the local X-ray morphology.

For the comparison with the new Chandra data we shall not present the
details of our model. They have been already described in the above paper
(Berezhko et al. \cite{bkv02}). We shall rather use some
simple analytical approximations,
needed to qualitatively interpret the spatial distribution of the various
CR components in SNRs. For the quantitative comparison we
shall however use the exact numerical results. With some additions, the
distribution of CRs produced by the spherically expanding shock of radius
$R_{\mathrm s}$ and speed $V_{\mathrm s}$ can be roughly described by the steady state 
1-dimensional transport equation for the CR distribution function 
$f(r,p,t)$
\begin{equation}
 \kappa \frac{\partial^2 f}{\partial x^2}
-u\frac{\partial f}{\partial x}
+\frac{p}{3}\frac{du}{dx}\frac{\partial f}{\partial p}
-L=0, \label{eq2}
\end{equation}
where $x=R_{\mathrm s}-r$, and $u=V_{\mathrm s}-w$ is the speed of the scattering medium
relative with respect to the shock front, $w$ is the speed in the frame of
the progenitor star, $\kappa(p)$ is the CR diffusion coefficient, $p$ is
particle momentum, $L$ is a loss term, which we take in the simple form
$L=f/\tau$, where $\tau$ is the loss time. Within this approach we neglect
any effects of the shock modification due to the CR backreaction.
Therefore the velocity profile has the form: $u(x<0)=u_1 = V_{\mathrm s};
\hspace{0.5cm} u(x>0)=u_2=u_1/\sigma$, where $\sigma = 4$ is the shock
compression ratio.
The solution of this transport equation is 
$f_i=f_0(p)\exp(-|x|/l_i)$ (V\"olk et al. \cite{vmf81}),
where $f_0(p)=f(x=0,p)$ is the CR distribution function at the shock
front and the scale 
\begin{equation}
l_i=
{2\kappa_i \over u_i} \left[
1-(-1)^i\sqrt{1+4\kappa_i/(u_i^2\tau_i)}
\right]^{-1}
\label{eq3}
\end{equation}
describes the spatial CR distribution in the upstream ($i=1$) and
downstream ($i=2$) regions, with simple limits for strong ($\tau_i\ll
\kappa_i/u_i^2$) and weak ($\tau_i\gg \kappa_i/u_i^2$) losses.
The CR distribution function at the shock front is determined by the
expression
\begin{equation}
f_0=Ap^{-q}\exp\left[
-\int _{p_{\mathrm inj}}^p dp \phi(p)/p\right],
\label{eq4}
\end{equation}
where $q=3u_1/(u_1-u_2)$;
$\phi=q[\kappa_1/(u_1^2\tau_1)+\kappa_2/(u_1u_2\tau_2)]$.

The losses produce two effects.  First of all, the universal power law
spectrum of accelerated CRs $f\propto p^{-q}$ has an exponential cutoff,
where the maximum CR momentum $p_{\mathrm m}$ is determined through the
condition $\phi(p_{\mathrm m})=1$. The losses lead also to a reduction of the
spatial scales $l_i$ that is essential in the cutoff region $p\sim
p_{\mathrm m}$, cf. eq.\ref{eq4}.

In the collisionless cosmic plasma, losses of nuclear CRs due to their
interactions with the gas particles or with the ambient fields are
negligibly small. The only important effects which restrict the proton
acceleration are adiabatic cooling in the downstream region and a
geometrical factor, that is the finite shock size. Formally compared with
the simplified one-dimensional eq.\ref{eq2}, the spherically symmetric
transport equation contains the additional term $(2\kappa_1/r)(\partial
f/\partial r)$ which contains the information about the shock size $R_{\mathrm s}$.
Approximately we can write $\partial f/\partial r \approx -f/l_1$.
Therefore this additional term in the plane wave approach can be expressed
as a loss term $L$ with the loss time $\tau_1= R_{\mathrm s}/(2u_1)$. In the
downstream region inside the SNR there is in addition also the term
$(\nabla \vec{u} p/3)(\partial f/\partial p)$ which describes CR adiabatic
cooling. It can be similarly estimated as a loss term with loss time
$\tau_2\sim R_{\mathrm s}/V_{\mathrm s}$. As a result, the upper proton momentum is determined
by the relation $\kappa_1(p_{\mathrm m})=R_{\mathrm s}V_{\mathrm s}/A$, where $A \sim 10$
(Berezhko \cite{ber96}).

In addition to the above mechanical effects CR electrons suffer
synchrotron losses with a time scale $\tau= 9 m_\mathrm{e}^2c^2/(4 r_0^2
B^2p)$, where $m_\mathrm{e}$ is the electron mass, $r_0$ denotes the
classical electron radius, and $B$ is the MF strength. This is
the dominant loss effect for our discussion. For sufficiently low MF,
when the loss time $\tau $ exceeds the age $t$ of the system,
synchrotron losses are not important and the electron and proton spectra
have exactly the same shape. In this case the highest energy electrons
with $p\sim p_{\mathrm m}$ have quite a wide spatial distribution with scales
$l_{1,2}\sim 0.1R_{\mathrm s}$. For high MF values the electron spectrum
is restricted to significantly lower momenta, $p^e_{\mathrm m}\ll p_{\mathrm m}$, than
are protons. In this case accelerated electrons occupy only a very thin
region around the shock front since $l_{1,2}\ll 0.1R_{\mathrm s}$ for all electron
energies, for which $\tau\ll t$. The radio-electrons in the GeV range have
$\tau\gg t$ and therefore $l_2\sim 0.1 R_{\mathrm s}$, consistent with observation
(Long et al. \cite{long}). The same is true for $100$~TeV electrons which
produce X-ray emission in the low-field case.

During the free expansion phase, preceding the early Sedov phase in which
SN~1006 is at the present time, a Rayleigh-Taylor instability of the
interface between the shocked interstellar gas and the ejected gas can
occur (Gull \cite{gull73}). It can make the downstream gas velocity
profile $w(r)$ narrower than the classical Sedov profile, leading to an
increase of the adiabatic expansion compared with the above estimate.  As
a result the radio emission profile may become somewhat narrower, whereas
the X-ray emitting electrons are not affected, being dominated by the
strong synchrotron losses around the shock This instability can also
locally amplify the MF. However, the unstable region moves to
the deep interior ($r > 0.1 R_{\mathrm s} $) as the Sedov phase sets in, and dies
out. Therefore we do not consider these processes any further. In
particular they have no influence on the CR acceleration which occurs near
the shock.
%
\begin{figure}
\centering
\includegraphics[width=7.2cm]{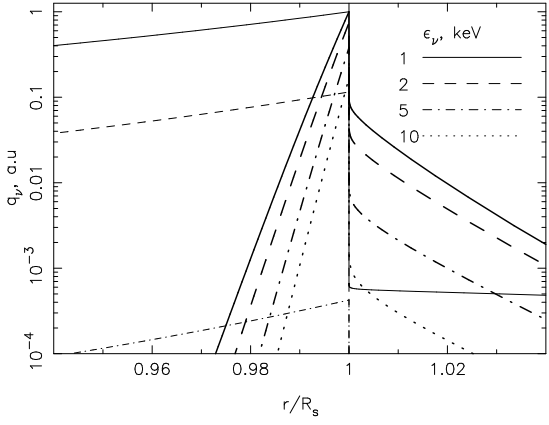}
\caption{Radial dependence of the X-ray emissivity at different
X-ray energies. Thick and thin lines correspond to the efficient and 
the so-called inefficient model respectively} \label{f1}
\end{figure}

Our nonlinear model (Berezhko et al.\ \cite{byk96}; Berezhko \& V\"olk
\cite{bv97}) is based on a fully time-dependent solution of the CR
transport equation together with the gas dynamic equations in spherical
symmetry. It yields at any given distance $r$ the momentum distribution
$f(r,p,t)$ of CRs produced during the SNR evolution up to time $t$.  
Because of the efficient acceleration of the nuclear CR component (mainly
protons), consistent with the expected injection rate of suprathermal
particles into the acceleration process (V\"olk et al.\ \cite{vbk03}), the
CRs significantly modify the shock structure by their nonlinear
backreaction and amplify the MF (Berezhko et al.
\cite{bkv02}). This MF amplification on all scales drives the CR
scattering mean free path towards the particle gyroradius $\rho_{\mathrm B}$, which
means isotropic spatial diffusion. Since the particle distribution is also
isotropic this implies the Bohm limit in the effective mean MF (V\"olk
\cite{voe84}): $\kappa=\rho_{\mathrm B} v/3$, where $v$ denotes the particle
velocity. Fig.\ref{f1} shows the calculated radial dependence of the
X-ray synchrotron emissivity $q_{\nu}(\epsilon_{\nu},r)$ (e.g. Berezinskii et
al.\ \cite{brz90})
%
%
for four X-ray energies $\epsilon_{\nu}=1$, 2, 5 and 10~keV. One can see
that the emissivity $q_{\nu}(r)$ peaks at the shock position $r=R_{\mathrm s}$
because the radiating electrons have a sharp peak at $r=R_{\mathrm s}$ due to the
synchrotron losses. Since within the upstream and downstream emission
regions the respective MF is approximately uniform, the
spatial behavior of the emissivity $q_{\nu}(r)$ is determined by the
spatial dependence of the CR distributions. According to eq.\ref{eq3},
the expected upstream scale for the energy
$\epsilon_{\nu}=5$~keV is $l_1=0.01R_{\mathrm s}$, taking into account that the
upstream MF is $B_1=\sigma_{\mathrm p} B_0$, the upstream medium speed
is $u_1=V_{\mathrm s}/\sigma_{\mathrm p}$, $\sigma_{\mathrm p}\approx 2$ is the precursor compression
ratio, $V_{\mathrm s}=3200$~km/s, $R_{\mathrm s}=7.3$~pc, and $B_0=20$~$\mu$G is the MF in
the shock precursor, i.e. the interstellar MF which is significantly
amplified by the (nonlinear) CR streaming instability. This is in good
agreement with the numerical results in Fig.\ref{f1}, based on a source
distance of 1.8 kpc.

The downstream scale due to the synchrotron losses, 
expected from relations (\ref{eq3}), is
%
$l_2\approx \sqrt{\kappa_2 \tau_2} \approx 0.0037 R_{\mathrm s}$, 
%
taking into account the values $u_2=V_{\mathrm s}/\sigma$, $B_2=\sigma B_0$, and
$\sigma=6.3$ (Berezhko et al. \cite{bkv02}). This agrees with 
the numerical values quite
well. Due to the energy-independence of $l_2$, the spectral form of the
downstream $q_{\nu}(\epsilon_{\nu}, r)$ does not depend on
$\epsilon_{\nu}$ (see Fig.\ref{f1}, in agreement with observations
(Bamba et al. \cite{bamba}).

The MF $B_1=B(r=R_{\mathrm s}+0)$, just ahead of the gas subshock at
$r=R_{\mathrm s}$, and the postshock MF $B_2=B(r=R_{\mathrm s}-0)$ are connected by the
relation $B_2=\sigma_{\mathrm s} B_1$, where $\sigma_{\mathrm s}=3.6$ is the subshock
compression ratio. Since the upstream MF is significantly smaller than
the downstream MF, the synchrotron emissivity undergoes a jump at
$r=R_{\mathrm s}$, and the emission from the downstream region significantly exceeds
the upstream emission.

In projection along the line of sight, the radial emissivity profile 
determines the remnant's surface brightness.
For the X-ray energy interval
$\epsilon_1<\epsilon_{\nu}<\epsilon_2$ it has the form
\begin{equation}
J_{\nu}(\rho)\propto 
\int _{\epsilon_1}^{\epsilon_2}d\epsilon_{\nu}
\int dx q_{\nu}(\epsilon_{\nu},r=\sqrt{\rho^2+x^2}) \label{eq6}
\end{equation}
where $\rho$ is the distance between the center of the remnant 
and the line of sight.
It is clear from this expression that, due to the shock curvature, the
surface brightness profile $J_{\nu}(\rho)$ differs
from the emissivity profile $q_{\nu}(r=\rho)$, except
for the simple case of a plane shock which is parallel to the line of
sight. In particular, the position $\rho_{\mathrm m} < R_{\mathrm s}$ of the peak value
$J_{\mathrm m}=J_{\nu}(\rho_{\mathrm m})$ does not coincide with the shock edge
$\rho=R_{\mathrm s}$, and the scaling values $L_{1,2}$ which characterize the
spatial brightness behavior in the inner $(\rho<\rho_{\mathrm m})$ and the outer
$(\rho>\rho_{\mathrm m})$ regions are not simply the downstream and upstream
emissivity scales $l_{1,2}$.

\begin{figure}
\centering
\includegraphics[width=7.2cm]{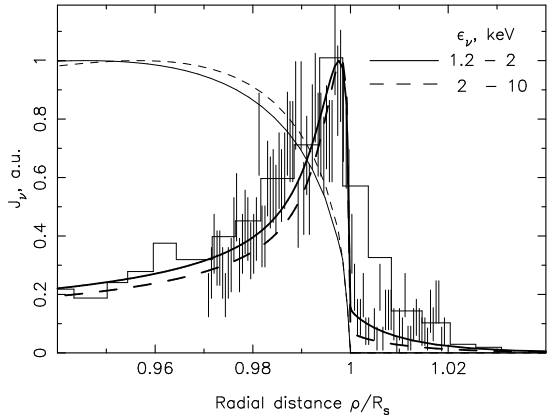}
\caption{Projected radial dependence of the X-ray brightness in the 
1.2 to 2 keV (solid) and 2 to 10~keV (dashed)
X-ray energy interval. Thick and thin lines correspond to the efficient and 
the so-called inefficient model respectively. The Chandra data, 
corresponding to the 
sharpest profile, are shown by the histogram 
(Long et al. \cite{bamba})
and the vertical dashes (Bamba et al. \cite{long}).} 
\label{f2}
\end{figure}

The numerically calculated brightness profile for the X-ray energy
interval between
$\epsilon_1=2$~keV and $\epsilon_2=10$~keV is shown in Fig.\ref{f2}.  
The brightness profile is characterized by the outer scale
$L_1=0.002R_{\mathrm s}=0.015$~pc which comes from the emission of the downstream
region alone, and from the shock curvature. The inner brightness scale is
$L_2\approx 7L_1=0.1$~pc.

The sharpest experimental X-ray brightness profiles obtained by the
Chandra observers (Long et al. \cite{long}; Bamba et al. \cite{bamba})
are shown
in Fig.\ref{f2}. Since the absolute values of the measurements are not 
known,
all theoretical and experimental profiles are normalized to their peak 
values.
One can see that the experimental values agree very well
with our calculations. All other observed brightness profiles are
significantly wider: on average $L_1=0.04$~pc and $L_2=0.2$~pc 
(Bamba et al. \cite{bamba}).
There are several reasons for a broadening of the observed profile,
compared with the ideal case of a spherical shell. First of all, it is
clear from the observations that the actual shock front deviates from a
spherical form. The wavy shape of the shock front can be produced as the
result of a small scale density inhomogeneity of the ambient interstellar
medium. Any small scale distortion of the spherical emission shell leads
to a broadening of the observed brightness profile.

It is important to note that there is direct experimental evidence that
not only the inner part $(\rho<\rho_{\mathrm m})$ of the brightness distribution
$J_{\nu}(\rho)$ but also the outer part $(\rho >\rho_{\mathrm m})$ is due to
emission from the downstream region $r<R_{\mathrm s}$, in contrast to the arguments
of Bamba et al. (\cite{bamba}). At the gas subshock the density increases
by a factor $\sigma_{\mathrm s} = 3.6$ and the gas temperature increases by a factor
of 9.3. Due to this fact the thermal emission from the upstream region is
small compared to the downstream thermal emission. Therefore, if the
observed nonthermal X-ray emission $J_{\nu}(\rho)$ at $\rho>\rho_{\mathrm m}$ is
due to the upstream electrons, then one should expect that the thermal
X-ray radiation has a peak value at a smaller distance than $\rho_{\mathrm m}$
and drops to almost zero at $\rho =\rho_{\mathrm m}$, if as before
$\rho_{\mathrm m}$ represents the peak position of the nonthermal X-ray emission
(in other words the peak of the thermal X-rays is expected in this case at
smaller distances).  However, the observed peak positions of the thermal
and nonthermal X-ray emissions and their shapes in the outer region
$\rho>\rho_{\mathrm m}$ are almost identical (Bamba et al. \cite{bamba}). This
is a strict confirmation that the entire observed X-ray emission comes
from the downstream region, exactly as predicted by Berezhko et al.
(\cite{bkv02}).

In Figs.~\ref{f1} and \ref{f2} we also present calculations for the
so-called inefficient model. Of course, the number of accelerated
electrons was chosen to be consistent with the observed synchrotron
emission for the assumed downstream MF $B_{\mathrm d}=16$~$\mu$G (see
Berezhko et al. \cite{bkv02} for details). Since in this case the MF
is much lower, so that synchrotron losses play no role, the upstream
(negligible brightness) and downstream electron length scales $l_{1,2}\sim
0.1R_{\mathrm s}$ are so large (see Fig.\ref{f1}) that the brightness profile is
almost two orders of magnitude wider than observed. The clear conclusion
is that inefficient scenarios for CR injection/acceleration in SN~1006
should be rejected since they strongly contradict the X-ray measurements.
%

Note that in order to form such a sharp decrease of their distribution
towards the center of the remnant one needs a powerful loss process in the
downstream region, independently of what specific acceleration process
produces the energetic electrons near the SN shock. Adiabatic cooling, as
it was demonstrated, is not strong enough. In the diluted ionized plasma
the only known strong electron loss process is synchrotron losses. But
this process can influence the 100~TeV electron distribution in a
significant form only if the downstream MF is as high as
$B_{\mathrm d}\sim 100$~$\mu$G, much larger than the typical interstellar MF.
Such a high MF can be created due to CR backreaction upstream of the
shock. 
For such a large downstream MF $B_{\mathrm d}\sim 100$~$\mu$G the number of
relativistic electrons consistent with the observed radio flux from
SN~1006 is so low (Berezhko et al. \cite{bkv02}) that they are not able to
create Alfv\'{e}n waves with sufficiently high amplitudes $\delta B\sim
B$.  Indeed, taking into account that a typical value of the Alfv\'{e}n
speed is $c_{\mathrm a}=20$~km/s and that for $B_{\mathrm d}\sim 100$~$\mu$G the pressure of
the CR electron component $P_{\mathrm c1}\approx 10^{-3}\rho_{\mathrm g} V_{\mathrm s}^2$ (Berezhko et
al. \cite{bkv02}), we have $(\delta B/B)^2\approx 0.1$ from eq.\ref{eq1},
which implies an insignificant background MF amplification. The only
possibility is efficient proton acceleration. In this case their number,
consistent with all existing data, is so high, $P_{\mathrm c}\sim \rho_{\mathrm g} V_{\mathrm s}^2$,
that they are able to strongly amplify the MF and at the same
time to provide efficient CR scattering approaching the Bohm limit (Bell
\& Lucek \cite{bluc01}).
Such a MF amplification is required by the comparison of our
selfconsistent model with the synchrotron observations. We note that the
uniformity of the downstream MF strength distribution during
the Sedov phase, despite the strong expansion of the downstream region, is
a direct consequence of the temporal weakening of the random MF
generation at the shock $B_2\propto V_{\mathrm s}^2$ (Bell \& Lucek \cite{bluc01}),
if we can approximate the MF as being isotropic according to the
relation $B_{\mathrm d}^2\rho_{\mathrm g}^{-4/3}= const$ (e.g. Chevalier \cite{chev74}).

We conclude that these data confirm the MF amplification and the
efficient acceleration of nuclear CRs in SN~1006, predicted by Berezhko
et al. (\cite{bkv02}). This efficiency is consistent with the requirements for
the Galactic CR energy budget.

This work has been supported in part by the Russian Foundation for Basic
Research (grant 03-02-16325). EGB and LTK acknowledge the
hospitality of the Max-Planck-Institut f\"ur Kernphysik, where part of
this work was carried out. 
LTK acknowledges the receipt of a JSPS
Research Fellowship.


\begin{thebibliography}{99}

\bibitem[2001]{apg}
Allen, G. E., Petre, R. \&  Gotthelf, E. V. 2001, ApJ, 558, 739

\bibitem[2003]{bamba}
Bamba, A., Yamazaki, R., Ueno, M., Koyama, K.\ 2003, ApJ, 589, 827

\bibitem[1978]{bell78}
Bell, A. R. 1978, MNRAS, 182, 147

\bibitem[2001]{bluc01}
Bell, A. R. \& Lucek, S. G. 2001, MNRAS, 327, 433

\bibitem[1996]{ber96}
Berezhko, E. G. 1996, Astropart. Phys., 5, 367

\bibitem[1996]{byk96} 
Berezhko, E. G., Elshin V. K. \& Ksenofontov, L. T. 1996, JETP, 82, 1

\bibitem[1997]{bv97}
Berezhko, E. G. \& V\"olk, H. J. 1997, Astropart. Phys., 7, 183

\bibitem[2002]{bkv02}
Berezhko, E.\ G., Ksenofontov, L.\ T., V\"olk, H.\ J.\ 2002, A\&A, 395, 943

\bibitem[1990]{brz90}
Berezinskii, V. S., Bulanov, S. A., Dogiel, V. A.,
Ginzburg, V.L. \& Ptuskin, V. S.
1990, Astrophysics of cosmic rays, ed. Ginzburg, V. L.
North-Holland, Amsterdam, 1

\bibitem[1978]{bo78}
Blandford, R. D. \& Ostriker, J. P. 1978, ApJ, 221, L29

\bibitem[1974]{chev74}
Chevalier, R.A. 1974, ApJ, 188, 501

\bibitem[1973]{gull73}
Gull, S.F. 1973, MNRAS, 161, 47

\bibitem[1995]{koyama}
Koyama, K., Petre, R., Gotthelf, E. V., et al. 1995, Nature, 378, 255 

\bibitem[2003]{long}
Long, K.S., Reynolds, S.P., Raymond, J.C., et al.\ 2003, ApJ, 586, 1162

\bibitem[2000]{lucb00}
Lucek, S. G. \& Bell, A. R. 2000, MNRAS, 314, 65

\bibitem[1982]{mckv82}
McKenzie, J. F. \& V\"olk, H. J. 1982, A\&A, 116, 191

\bibitem[1998]{tan98}
Tanimori, T., Hayami, Y., Kamei, S., et al. 1998, ApJ, 497, L25

\bibitem[1984]{voe84}
V\"olk, H. J. 1984, High Energy Astrophysics, ed. J. Tran Thanh Van,
Editions Frontieres, Gif sur Yvette.

\bibitem[2003]{vbk03}
V\"olk, H.J., Berezhko, E.G.  \& Ksenofontov, L.T. 2003, A\&A, 409,
563

\bibitem[1981]{vmf81}
V\"olk, H. J., Morfill, G. E. \& Forman, M. A. 1981, ApJ, 249, 161

\bibitem[1997]{wink}
Winkler, P. F. \& Long, K. S. 1997, ApJ, 491, 829

\end{thebibliography}
\end{document}